\documentclass[aps,prc,twocolumn,floatfix,nofootinbib,showpacs,superscriptaddress,tightenlines]{revtex4-1}
\usepackage{amsmath}
\usepackage{amssymb}
\usepackage{amsthm}
\usepackage{dcolumn}
\usepackage{epsfig}
\usepackage{graphics}
\usepackage{graphicx}
\usepackage{amsmath}
\usepackage{longtable}
\usepackage{color}

\definecolor{darkgreen}{rgb}{0,0.5,0}
\definecolor{purple}{rgb}{0.5,0,0.5}
\definecolor{nblue}{rgb}{0.0,0.0,0.50}
\definecolor{scarlet}{rgb}{1.0,0.2,0}

\newcommand{\nslash}{\mbox{$\not \! n$}}
\newcommand{\ellslash}{\mbox{$\not \! \ell$}}
\newcommand{\qslash}{\mbox{$\not \! q$}}
\newcommand{\Pslash}{\mbox{$\not \! P$}}

\begin{document}

\title{Pion and kaon valence-quark parton distribution functions}

\author{Trang Nguyen}\affiliation{Center for Nuclear Research, Department of
Physics, Kent State University, Kent OH 44242, USA}

\author{Adnan Bashir}
\affiliation{Instituto de F\'{\i}sica y Matem\'aticas,
Universidad Michoacana de San Nicol\'as de Hidalgo, Apartado Postal
2-82, Morelia, Michoac\'an 58040, Mexico}
\affiliation{Kavli Institute for Theoretical Physics China, CAS, Beijing 100190, China}

\author{Craig~D.~Roberts}
\affiliation{Kavli Institute for Theoretical Physics China, CAS, Beijing 100190, China}
\affiliation{Physics Division, Argonne National
Laboratory, Argonne, Illinois 60439, USA}
\affiliation{Department of Physics, Center for High Energy Physics and the State Key Laboratory of Nuclear Physics and Technology, Peking University, Beijing 100871, China}

\author{Peter~C.~Tandy}
\affiliation{Center for Nuclear Research, Department of
Physics, Kent State University, Kent OH 44242, USA}
\affiliation{Kavli Institute for Theoretical Physics China, CAS, Beijing 100190, China}

\begin{abstract}
A rainbow-ladder truncation of QCD's Dyson-Schwinger equations, constrained by existing applications to hadron physics, is employed to compute the valence-quark parton distribution functions of the pion and kaon.  Comparison is made to $\pi$-$N$ Drell-Yan data for the pion's $u$-quark distribution and to Drell-Yan data for the ratio $u_K(x)/u_\pi(x)$: the environmental influence of this quantity is a parameter-free prediction, which agrees well with existing data.  Our analysis unifies the computation of distribution functions with that of numerous other properties of pseudoscalar mesons.
\end{abstract}


\maketitle

\date{\today}
Experimental information on the quark and gluon parton distribution functions (PDFs) in the pion have primarily been inferred from the Drell-Yan reaction \cite{Badier:1983mj,Betev:1985pg,Conway:1989fs} in pion-nucleon and pion-nucleus collisions.  Kaon PDF data exists in the form of the ratio $u_K(x)/u_\pi(x)$ \cite{Badier:1980jq,Badier:1983mj}.  While the nucleon PDFs are now fairly well determined, the pion and kaon PDFs remain poorly known.   Reference \cite{Holt:2010vj} reviews both the experimental and theoretical status of nucleon and pion PDFs.  Since the pion is central to hadron physics, and its key characteristics are dictated by dynamical chiral symmetry breaking, pion structure is a critical testing ground for our understanding of nonperturbative QCD.   Much more theoretical work has been devoted to the pion elastic charge form factor (e.g., \cite{Maris:2000sk}); $\pi\pi$ scattering (e.g., \cite{Bicudo:2001jq}); and the pion electromagnetic transition form factor (e.g., \cite{Roberts:2010rn}) than has been devoted to the pion PDFs.  Herein we take a material step toward ameliorating that deficit.

Numerical simulations of lattice-regularized QCD are restricted to the computation of low-order moments of the PDFs: the pointwise dependence is not directly accessible \cite{Holt:2010vj,Best:1997qp}.  Model calculations of PDFs are challenging because the framework employed must necessarily possess perturbative QCD features in coexistence with the covariant, nonperturbative description of a bound state.  Chiral symmetry has guided studies of pion PDFs within the Nambu--Jona-Lasinio (NJL) model \cite{Shigetani:1993dx,Weigel:1999pc} at the expense of: an unphysical point-particle structure for the pion Bethe-Salpeter amplitude; and ambiguities from  a dependence upon regularization procedure owing to the lack of renormalizability.    Constituent quark models \cite{Szczepaniak:1993uq} and  instanton-liquid models \cite{Dorokhov:2000gu} have also been used.
In all these approaches, it is difficult to have pQCD elements join smoothly with nonperturbative aspects while respecting the quantum field theoretical nature of the underlying dynamics.
The large $x$ behavior of the pion PDFs provides an illustration.   The QCD parton model \cite{Farrar:1975yb} and pQCD \cite{Brodsky:1994kg} are clear: at a scale of order-$\Lambda_{\rm QCD}$ the behavior is $u_\pi(x)
\propto (1-x)^\alpha$ with \mbox{$ \alpha = 2 +\gamma$} where $\gamma>0$ is a logarithmic correction.   However the above models imply an $\alpha$ ranging from 0 to 1, or at most 1.5~\cite{Holt:2010vj}.

These issues may in principle be addressed if the PDFs can be obtained from truncations of QCD's Dyson-Schwinger equations.   The DSEs are a hierarchy of coupled integral equations for the Schwinger  functions ($n$-point functions) of a theory.  Bound-states appear as poles in the appropriate $n$-point functions; e.g., the bound-state Bethe-Salpeter  equation (BSE) of field theory appears after taking residues in the inhomogeneous DSE for the appropriate color singlet vertex.  Numerous reviews; e.g.,  \cite{Roberts:1994dr}, describe the insight into hadron physics achieved through the use of the rainbow-ladder (RL) truncation of the DSEs, which is the leading-order in a systematic, symmetry-preserving scheme \cite{Munczek:1994zz}.

The first DSE study of PDFs was conducted for the pion \cite{Hecht:2000xa} in an analysis that employed phenomenological parametrizations of both the Bethe-Salpeter amplitude and dressed-quark propagators.  The purpose of this present work is, for the first time: to employ numerical DSE solutions in the computation of the pion and kaon PDFs, adapting the RL model employed in successful predictions of electromagnetic form factors \cite{Maris:1999bh,Maris:2000sk,Maris:2002mz,Holl:2005vu}; and study the ratio $u_K(x)/u_\pi(x)$ in order to elucidate aspects of the influence of an hadronic environment.

In the Bjorken-limit, DIS selects the most singular behavior of a correlator of quark fields of the target with light-like and causal distance separation $z^2 \sim 0^+ $.   With incident photon momentum along the negative
3-axis, the kinematics selects \mbox{$z^+ \sim z_\perp \sim 0$}  leaving  $z^-$ as the finite distance conjugate to quark momentum component $xP^+$, where  \mbox{$ x = Q^2/2P \cdot q$} is the Bjorken variable, \mbox{$q^2 = -Q^2$} is the spacelike virtuality of the photon, and $P$ is the target momentum.  To leading  order in the operator product expansion, the target structure functions are proportional to the charge-weighted sum of PDFs, $q_f(x)$, for parton of flavor $f$.    The PDF is  given by the  correlator~\cite{Jaffe:1983hp,Ellis:1991qj}
\begin{equation}
q_f(x) = \frac{1}{4 \pi} \int d\lambda \, e^{-i x P\cdot n \lambda }
\langle \pi(P) | \bar{\psi}_f (\lambda n) \, \nslash
\, \psi_f(0) | \pi(P) \rangle_c ~~,
\label{Mink_dis_x_inv}
\end{equation}
expressed here in manifestly Lorentz-invariant form.  In the infinite momentum frame, $q_f(x)$ is the probability that a single $f$-parton has momentum fraction
\mbox{$x=k\cdot n/P\cdot n $} \cite{Ellis:1991qj}.  In the above, $n^\mu$, and (for later use) $p^\mu$, are  light-like  vectors satisfying \mbox{$n^2 = p^2 = 0$} and \mbox{$n \cdot p = 2$}.  They form a convenient basis for the longitudinal sector of 4-vectors.   One has \mbox{$k \cdot n = k^+$} and \mbox{$k \cdot p = k^-$}.    The dominant component of $q$ is parallel to $n$, i.e., $q^-$ dominates.  Note that \mbox{$q_f(x) = - q_{\bar{f}}(-x) $}, and that the valence quark amplitude is \mbox{$ q^v_f(x) = q_f(x) - q_{\bar{f}}(x)$}.

It follows from Eq.\,(\ref{Mink_dis_x_inv}) that \mbox{$ \int_0^1 dx \,  q^v_f(x) = $} \mbox{$ \langle \pi(P) | J^+_f(0) | \pi(P) \rangle /2P^+ =$}  \mbox{$  F_\pi(0) = 1$}.   Approximate treatments should at least preserve  vector current conservation to automatically obtain the correct normalization.

In our DSE framework, dynamical information on the various nonperturbative elements, such as propagators and bound state amplitudes, is available in a Euclidean momentum representation.   (In our Euclidean metric: $\{\gamma_\mu,\gamma_\nu\}
= 2\delta_{\mu\nu}$; $\gamma_\mu^\dagger = \gamma_\mu$; $\gamma_5=
\gamma_4\gamma_1\gamma_2\gamma_3$; $a \cdot b = \Sigma_{i=1}^4 a_i b_i$; $\not\!n = \gamma\cdot n$; and $P_\mu$ timelike $\Rightarrow$ $P^2<0$.)  The corresponding formulation of Eq.\,(\ref{Mink_dis_x_inv}) is
%
%
\begin{equation}
q_f(x) = -\frac{1}{2} \int \frac{d^4k}{(2\pi)^4}  \delta(k\cdot n - x P\cdot n) \; {\rm tr}_{\rm cd}
[ i\nslash \, G(k,P) ]~~~,
\label{Eucl_pdf_k}
\end{equation}
where ${\rm tr}_{\rm cd}$ denotes a color and Dirac trace, and $G(k,P)$ represents the forward $\bar q$-target scattering amplitude.  In Euclidean metric the vectors $n$, $p$, $P$ satisfy \mbox{$n^2=0=p^2 $}, \mbox{$n \cdot p = -2$}, \mbox{$P^2 = -m_\pi^2$},  and \mbox{$P \cdot n = - m_\pi$}.

\begin{figure}[t]
\includegraphics[clip,height=0.3\textwidth]{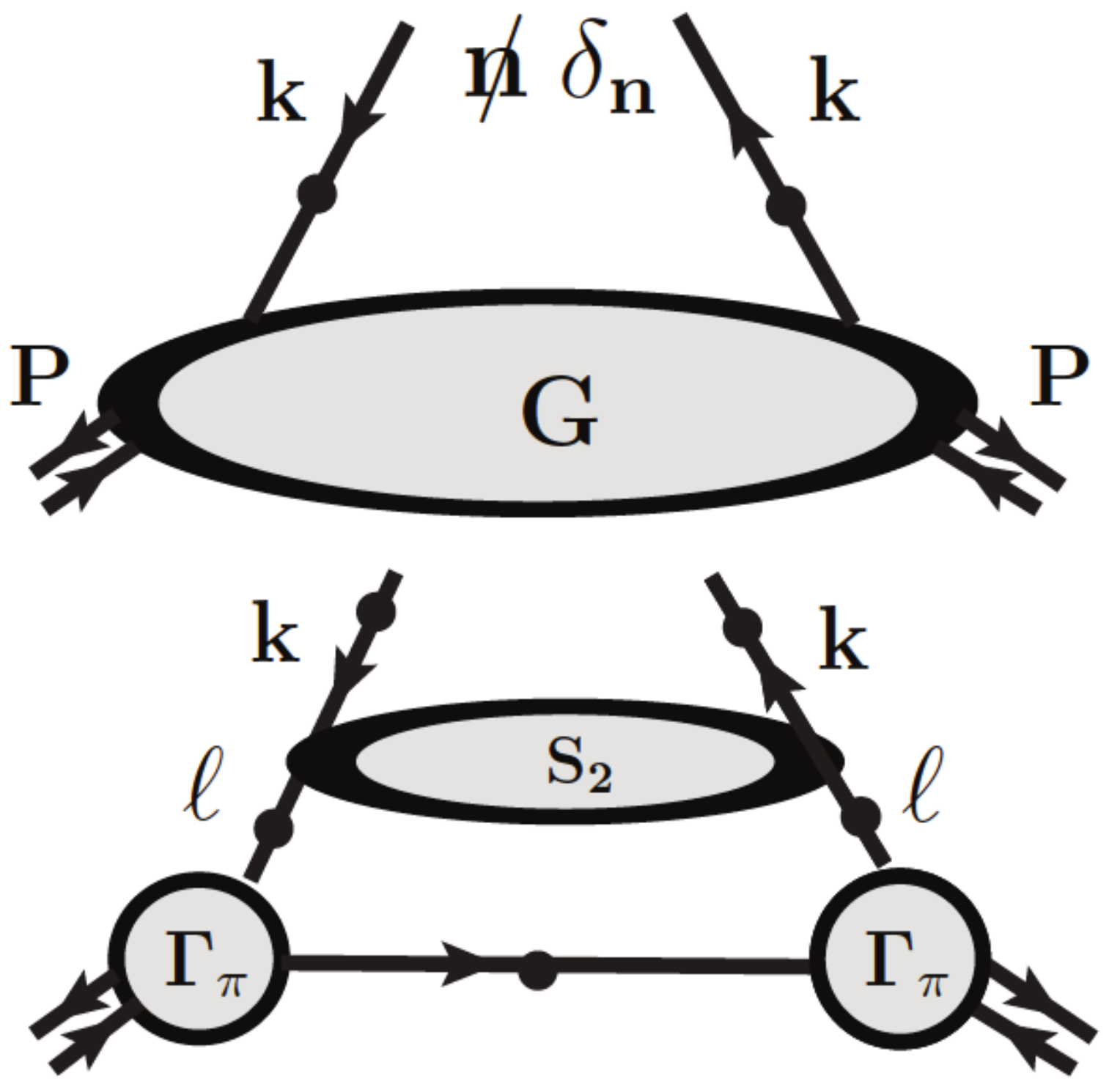}

\caption{ (Color online)  Diagrammatic representation of parton distributions.   \emph{Top panel} -- the exact parton distribution corresponding to Eq.~(\protect\ref{Eucl_pdf_k}); and \emph{bottom} -- Rainbow-ladder truncation of the amplitude $G$.    $S_2$ is the $q \bar q$ propagator.
\label{fig:PDF_exact_LR}}
\end{figure}

The top part of Fig.~\ref{fig:PDF_exact_LR} illustrates Eq.~(\ref{Eucl_pdf_k}).  In rainbow-ladder truncation, which treats only the valence $q \bar q$ structure of the target, we have the decomposition illustrated in the bottom part of Fig.~\ref{fig:PDF_exact_LR}.  The 4-point function $S_2$ is the $q\bar q$ two-body propagator and $\Gamma_\pi$ is the Bethe-Salpeter bound-state amplitude, both computed in the RL truncation.  The combination
\mbox{$S_2 \otimes \, i\nslash \, \delta(k\cdot n - x P\cdot n) = $}
\mbox{$S(\ell) \, \Gamma^n(\ell;x) \, S(\ell)$}, where $\Gamma^n(\ell;x)$ is a dressed vector vertex.  The RL truncation for the valence $u_\pi(x)$ is thus
\begin{eqnarray}
\nonumber
\lefteqn{u_\pi(x) = -\frac{1}{2} \int \frac{d^4 \ell}{(2\pi)^4}  {\rm tr}_{\rm cd}\, \left[ \Gamma_\pi(\ell,P)\right.}\\
&& \left. \times \,S_u(\ell)\, \Gamma^n(\ell;x) \, S_u(\ell)\, \Gamma_\pi(\ell,P)\, S_d(\ell-P) \right]\,,
\label{Eucl_pdf_LR_Ward}
\end{eqnarray}
wherein $\Gamma^n(\ell;x)$ is a generalization of the dressed-quark-photon vertex, describing a dressed-quark scattering from a zero momentum photon.  It satisfies the usual inhomogeneous BSE (here with a RL kernel) except that the inhomogeneous term is  $i\nslash \, \delta(\ell \cdot n - x P\cdot n)$.  The dressed-quark propagator is \mbox{$S(\ell;\zeta) = $}
\mbox{$ 1/[i \ellslash \, A(\ell^2;\zeta) + B(\ell^2;\zeta)] $}, where $\zeta$ is the renormalization mass scale.

This selection of RL dynamics parallels precisely the symmetry-preserving treatment of the pion charge form factor at \mbox{$Q^2 = 0 $}, wherein the vector current is conserved by use of ladder dynamics at all three vertices and rainbow dynamics for all three quark propagators \cite{Maris:2000sk,Maris:1999bh}.   Equation~(\ref{Eucl_pdf_LR_Ward}) ensures  \mbox{$ \int_0^1 dx \,  q^v_f(x) = 1$} for \mbox{$f = u, \bar d$} automatically  since $ \int dx \, \Gamma^n(\ell;x)$ gives the Ward-identity vertex and the result follows from canonical  normalization of the BS amplitude.

We adopt the representation \mbox{$\ell^\mu = \frac{1}{2} (\alpha p^\mu + \beta n^\mu)  + k_\perp^\mu $} to transform to new variables \mbox{$\alpha = $} \mbox{$- \ell \cdot n$} and \mbox{$\beta =$} \mbox{$ - \ell \cdot p$}, thus converting  Eq.~(\ref{Eucl_pdf_LR_Ward}) to the form
%
%
\begin{equation}
u_\pi(x) = \frac{-J_{\rm E}}{2(2\pi)^4}  \, \int_{-\infty}^{+\infty} d \beta \; d^2\ell_\perp \,
T(n, p; \ell, P)\big |_{\alpha = x P\cdot n},
\label{Eucl-Mink_pdf_k_LR}
\end{equation}
where: $J_{\rm E}=-i/2$ is the Jacobian of the variable transformation; and $T$ is the result of the trace in Eq.~(\ref{Eucl_pdf_LR_Ward}), using \mbox{$\Gamma^n(\ell;x) \approx $} \mbox{$ n_\mu \partial S^{-1}(\ell)/\partial \ell_\mu \, \delta(\ell \cdot n - x P\cdot n)$}, which is the correct result from the Ward Identity after $\int dx $.

Since $q_f(x)$ is obtained from the hadron tensor $W^{\mu \nu}$, which in turn can be formulated from the discontinuity  \mbox{$T^{\mu\nu}(\epsilon) -  T^{\mu\nu}(-\epsilon)$} of the forward Compton amplitude $T^{\mu\nu}$, we observe that all enclosed singularities from the difference of contours cancel except for the cut that produced the delta function constraint on $\alpha$.

We employ the RL-DSE model developed in Refs.\,\cite{Maris:1997hd,Maris:1997tm,Maris:1999nt}, in which the BSE kernel takes the form \mbox{$K  = $}  \mbox{$ -4\pi\,\alpha_{\rm eff}(k^2)\, D_{\mu\nu}^{\rm free}(k)
\textstyle{\frac{\lambda^i}{2}}\gamma_\mu \otimes \textstyle{\frac{\lambda^i}{2}}\gamma_\nu $},
where $k$ is the gluon momentum.  Here $\alpha_{\rm eff}(k^2)$ is a model running
coupling chosen such that it reproduces QCD's one-loop renormalization-group behavior for $k^2\gtrsim 2\,$GeV$^2$.  A more general method for treating $K$ has recently become available~\cite{Chang:2009zb}.  The DSE that produces the dressed quark propagator is also determined by $\alpha_{\rm eff}(k^2)$ \cite{Maris:1997hd,Maris:1997tm,Maris:1999nt};
and the combination of the DSE and BSE produces dressed color-singlet vector and axial-vector vertices satisfying their respective Ward-Takahashi identities.  This ensures that the chiral-limit ground-state pseudoscalar bound-states are the massless Goldstone bosons from dynamical
chiral symmetry breaking~\cite{Maris:1997hd,Maris:1997tm}; and it
ensures electromagnetic current conservation~\cite{Roberts:1994hh}.
This kernel is found to be successful for, amongst other things, light-quark meson  properties \cite{Maris:1999nt} including  electromagnetic elastic \cite{Maris:1999bh,Maris:2000sk} and transition \cite{Maris:2002mz,Holl:2005vu} form factors.   The model parameters are the two current quark masses and one infrared strength for $\alpha_{\rm eff}(k^2)$.    Selected results related to the pion and kaon are displayed in Table~\ref{tab:LR-MT_results}.
%

\begin{table}[t]
\caption{Illustrative selection of DSE results \protect\cite{Maris:1999nt,Maris:2000sk,Maris:1999bh,Maris:2002mz} obtained with the RL kernel employed herein compared with experimental values \protect\cite{PDG10}.
(Dimensioned quantities are listed in GeV or fm$^2$, as appropriate.)
\label{tab:LR-MT_results}
}
\begin{center}
\begin{tabular*}
{\hsize}
{
l@{\extracolsep{0ptplus1fil}}
|c@{\extracolsep{0ptplus1fil}}
c@{\extracolsep{0ptplus1fil}}
|c@{\extracolsep{0ptplus1fil}}
c@{\extracolsep{0ptplus1fil}}
|c@{\extracolsep{0ptplus1fil}}
c@{\extracolsep{0ptplus1fil}}
|c@{\extracolsep{0ptplus1fil}}
c@{\extracolsep{0ptplus1fil}}
}
 & $m_\pi$ & $f_\pi$ & $m_K$ & $f_K$ & $r_\pi^2$ &  $r_{K^+}^2$ & $g_{\pi \gamma \gamma}$ & $r_{\pi\gamma\gamma}^2$
 \\\hline
expt.  & 0.138 & 0.092 & 0.496   & 0.113 & 0.44 & 0.34 & 0.5 & 0.42 \\
calc. & 0.138 & 0.092 & 0.497   & 0.110 & 0.45 & 0.38 & 0.5 & 0.41 \\
\end{tabular*}
\end{center}
\end{table}

In the evaluation of Eq.\,(\ref{Eucl-Mink_pdf_k_LR}) we employ the full pseudoscalar meson  Bethe-Salpeter amplitude
\begin{eqnarray}
\nonumber
\lefteqn{\Gamma_{\pi}(\ell,P) = \gamma_5 \left[ i E_\pi(q;P) + \Pslash F_\pi(q;P) \right.}\\
&& \left. + \qslash \, G_\pi(q;P) + \sigma_{\mu\nu} q_\mu P_\nu H_\pi(q;P) \right]~,
\label{gen_Gamma_pi}
\end{eqnarray}
where \mbox{$q = \ell - P/2 $} is the relative $q \bar q$ momentum appropriate to
Eq.~(\ref{Eucl_pdf_LR_Ward}).    For a charge-conjugation eigenstate (e.g., the pion), the invariant amplitudes $E, F$ and $H$ are even in $q\cdot P$, while $G$ is odd.  The kaon invariant amplitudes contain both even and odd components.   We expand the $q\cdot P$ dependence in Chebschev polynomials \cite{Maris:1999nt}, keeping terms of order \mbox{$ n=0-3$}.   The domain of $\ell^2$ over which the quark propagators are needed in this application is larger  than what is available from previous solutions of the quark DSE.  We therefore adopt a constituent mass pole approximation for the denominator of the spectator quark propagator \cite{Hecht:2000xa}.    A constituent spectator mass of $0.4$\,GeV permits a minimal adjustment to establish the normalization $\langle x^0 \rangle$.
We compared the approximation \mbox{$\Gamma^n(\ell;x) \approx $} \mbox{$ n_\mu \partial S^{-1}(\ell)/\partial \ell_\mu \, \delta(\ell \cdot n - x P\cdot n)$} with the bare vertex truncation and found that both give essentially the same distributions after re-enforcing the normalization \mbox{$\langle x^0 \rangle = 1$}.
%

%
\begin{figure}[t]
\includegraphics[clip,width=0.5\textwidth]{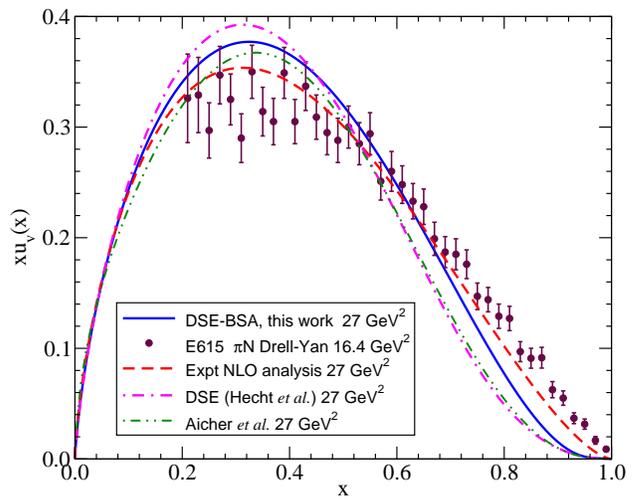}

\caption{(Color online) Pion valence quark distribution function evolved to (5.2~GeV)$^2$.  \emph{Solid curve} -- full DSE calculation \protect\cite{Nguyen_PhD10}; \emph{dot-dashed curve} -- semi-phenomenological DSE-based calculation in Ref.\,\protect\cite{Hecht:2000xa}; \emph{filled circles} -- experimental data from Ref.\,\protect\cite{Conway:1989fs}, at scale (4.05\,{\rm GeV})$^2$;
\emph{dashed curve} -- NLO re-analysis of the experimental data \protect\cite{Wijesooriya:2005ir};
and \emph{dot-dot-dashed curve} -- NLO reanalysis of experimental data with inclusion of soft-gluon resummation \protect\cite{Aicher:2010cb}. \label{fig:pi_DSE}}
\end{figure}

In Fig.~\ref{fig:pi_DSE} we display our DSE result \cite{Nguyen_PhD10} for the valence $u$-quark distribution evolved to $Q^2 = (5.2~{\rm GeV})^2$ in comparison with $\pi N$ Drell-Yan data \cite{Conway:1989fs} at a scale  $Q^2 \sim (4.05~{\rm GeV})^2$ obtained via a LO analysis.  Our distribution at the model scale $Q_0$ is evolved using leading-order DGLAP.    The model scale is fixed to \mbox{$Q_0 = 0.57 $}~GeV by matching the $x^n$ moments for $n=1,2,3$ to the experimental analysis given at (2\,{\rm GeV})$^2$ \cite{Sutton:1991ay}.  Our momentum sum rule result \mbox{$2 \, \langle x \rangle = $} 0.74~(pion), 0.76~(kaon) at $Q_0$ shows clearly the implicit inclusion of gluons as a dynamical entity in a true covariant bound-state approach.

In Fig.\,\ref{fig:pi_DSE} we also show the result from the first DSE study \cite{Hecht:2000xa}, which employed phenomenological parametrizations of the nonperturbative elements.  Our present calculation lies marginally closer to the Drell-Yan data in Ref.\,\cite{Conway:1989fs} at high-$x$.  However, this is not significant because both DSE results agree with pQCD; viz., \mbox{$u(x) \sim$}  \mbox{$ (1-x)^\alpha$} with  \mbox{$\alpha \gtrsim 2$} and growing with increasing scale, which is not true of the reported Drell-Yan data.

Motivated by this, a NLO reanalysis of the data was performed \cite{Wijesooriya:2005ir}; and we also show that result at  $Q^2 = (5.2~{\rm GeV})^2$ in Fig.~\ref{fig:pi_DSE}.   It does clearly reduce the extracted PDF at high-$x$ but not enough to resolve the data's apparent discrepency with pQCD behavior, which is discussed at length in  Ref.\,\cite{Holt:2010vj}.  The DSE exponents are  2.4 at  model scale \mbox{$Q_0 = 0.54$}\,GeV in Ref.\,\cite{Hecht:2000xa}, and 2.1 at scale \mbox{$Q_0 = 0.57$}\,GeV for the present study.  DSE analyses do not allow much room for a larger PDF at high-$x$.
A resolution of the conflict between data and well-constrained theory has recently been proposed: a reanalysis of the original data at NLO with a resummation of soft gluon processes \cite{Aicher:2010cb} produces a PDF whose behavior for \mbox{$x > 0.4$} is essentially identical to that of the earlier DSE calculation \cite{Hecht:2000xa}, as is apparent in Fig.\,\ref{fig:pi_DSE}.

%
\begin{figure}[t]
\includegraphics[clip,width=0.5\textwidth]{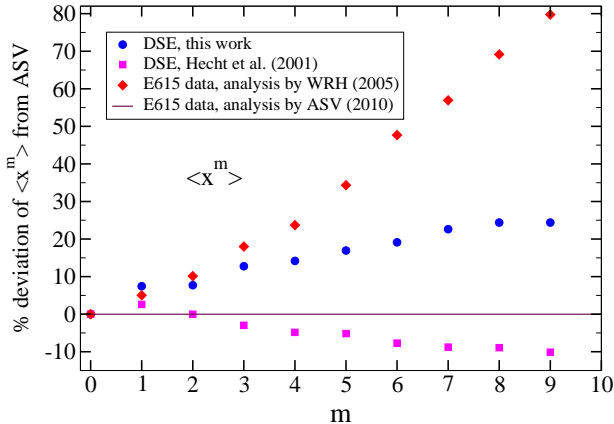}

\caption{(Color online)
Moments of the pion's valence $u(x)$ at scale  (5.2~GeV)$^2$, shown as a  \% deviation from the recent (ASV) re-analysis \protect\cite{Aicher:2010cb} (NLO, with soft gluon resummation) of the 1989 E615 $\pi N$ Drell-Yan data \protect\cite{Conway:1989fs}.  \emph{Filled circles} -- present full DSE calculation \protect\cite{Nguyen_PhD10}; \emph{filled squares} -- semi-phenomenological DSE-based calculation \protect\cite{Hecht:2000xa}; and \emph{filled diamonds} -- re-analysis (NLO, without soft gluon resummation) of the same Drell-Yan data \protect\cite{Wijesooriya:2005ir}.
\label{fig:pi_moments_5GeV}}
\end{figure}

In Fig.~\ref{fig:pi_moments_5GeV} we display the first nine moments of our result for  $u_\pi(x)$ at scale $Q^2 = (5.2~{\rm GeV})^2$ in comparison with the earlier DSE result  from Ref.~\cite{Hecht:2000xa} and the NLO reanalysis \cite{Wijesooriya:2005ir} of the original E615 data, all plotted as a \%-deviation from the moments of the most recent analysis of Ref.\,\cite{Aicher:2010cb}. Considering that the high moments are small, e.g., \mbox{$\langle x^9 \rangle \sim 0.003$}, the two DSE results are both equally well in accord with the recent analysis.

\begin{figure}[t]
\includegraphics[clip,width=0.5\textwidth]{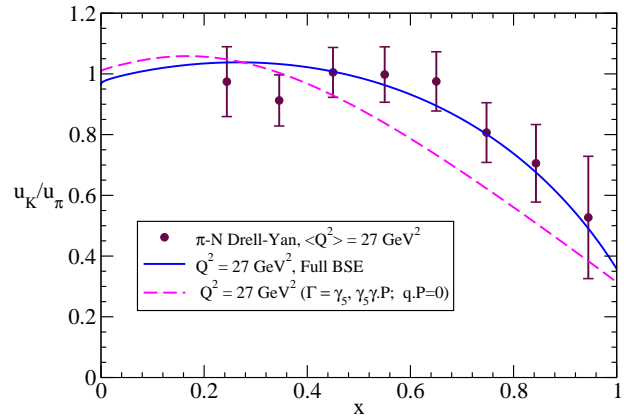}

\caption{(Color online) DSE prediction for the ratio of $u$-quark distributions in the kaon and pion \protect\cite{Nguyen_PhD10,Holt:2010vj}.   The full Bethe-Salpeter amplitude produces the \emph{solid} curve; the reduced BSE vertex produces the \emph{dashed} curve.  The reduced amplitude retains only the  invariants and amplitudes  involving  pseudoscalar and axial vector Dirac matrices, and ignores dependence on the variable $q\cdot P$.  These are part of the reductions that occur in a pointlike treatment of the pseudoscalar mesons.  The experimental data is from~\protect\cite{Badier:1980jq,Badier:1983mj}.
\label{fig:pi_DSE_ratio} }
\end{figure}

The ratio $u_K/u_\pi$ measures the effect of the local hadronic environment.   In the kaon, the $u$-quark is partnered with a  heavier partner than in the pion and this should cause $u(x)$ to peak at lower-$x$ in the kaon.  Our DSE calculation \cite{Nguyen_PhD10} is shown in Fig.\,\ref{fig:pi_DSE_ratio} along with available Drell-Yan data \cite{Badier:1980jq,Badier:1983mj}, which does not separate sea and valence quarks.  Our parameter-free result agrees well with the data.   It is notable that the value of this ratio at $x = 0$ approaches one under evolution owing to the increasingly large population of sea-quarks produced thereby \cite{Chang:2010xs}.  On the other hand, the value at $x=1$ is a fixed-point under evolution; i.e., it is the same at every value of the resolving scale $Q^2$, and is therefore a persistent probe of nonperturbative dynamics \cite{Holt:2010vj}.

In Fig.\,\ref{fig:pi_DSE_ratio} we  also display a calculation which employs a reduced BSE vertex: only the leading two invariant amplitudes  $E(q; P)$ and $F(q; P)$ are retained,  and each  is truncated to the lowest Chebychev moment in $q\cdot P$, i.e., \mbox{$E(q;P) \to \tilde{E}(q^2)  $}.   The field theory variable $q\cdot P$ is a constant in quantum mechanics.   (These reductions  in the BSE vertices occur within a NJL model description; but that model also ignores the $q^2$ dependence of the vertices.)    These simplifications do not change the qualitative behavior of the ratio, but the detailed quantitative agreement is impaired.

%

%
The ratio in Fig.\,\ref{fig:pi_DSE_ratio} was also computed using a nonpointlike-pion-regularized version of an NJL model \cite{Shigetani:1993dx}.  In this case one finds that while the behaviors of $u_v^\pi(x)$ and $u_v^K(x)$ are separately poor, the ratio is nevertheless in fair agreement with extant data: the result is similar to the dashed-curve in the figure.

An estimate of the leading large-$x$ behavior \mbox{$u_K(x) \sim A_K \, (1-x)^\alpha  $} can be made in the limit where the quark propagators are characterized  by constituent masses  $M_u, M_s$ and the vertex  is taken to be $i\nslash \, \delta(\ell\cdot n - xP\cdot n)$, preserving the Ward Identities.    We also truncate $\Gamma_\pi$  to
\mbox{$ \gamma_5 \,E_K( q^2) = $}  \mbox{$\gamma_5 \,N_K/( q^2 + \Lambda_K^2)  $} where
\mbox{$q = \ell - P/2  $}.    The quark mass dependence of $A_K$ and $A_\pi$ will provide an estimate of $u_K(1)/u_\pi(1)$.  For \mbox{$x > 1/2$} the pole in the spectator quark propagator is the only one in the upper half plane and the $\ell^-$ integral may readily be evaluated to yield
\begin{equation}
u_K(x) = \frac{4 N_c \pi^2}{(2\pi)^4} \int_{\mu_m(x)}^\infty d\mu \; \frac{x\, {\cal M}^2 + \mu + M_u^2}{[\mu + M_u^2]^2} \; E_K^2(q^2).
\label{NJL+mu_int}
\end{equation}
Here: \mbox{${\cal M}^2 = m_K^2 - (M_s -M_u)^2$}; we have changed the integration variable from $\vec{\ell}_\perp$ to \mbox{$ \mu = - \ell^2 $}, where the latter is the value at the $\ell^-$ pole; $q^2$ evaluated at the $\ell^-$ pole is \mbox{$ q^2 =$}  \mbox{$ m_K^2/4 + (\mu - M_s^2)/2$}; and \mbox{$\mu_m(x) =$}  \mbox{$ a/(1-x) - x m_K^2  $}, with \mbox{$a = x M_s^2  $}.   This divergence of the lower limit for large $x$ guarantees that the result is completely determined by the ultraviolet behavior of the propagators and bound state amplitudes.

The integral can be expressed as
\begin{equation}
u_K(x) = N \int_0^\infty d\hat{\mu} \; \frac{\frac{a}{1-x} + b + \hat \mu}{[\frac{a}{1-x} + c + \hat \mu ]^2} \; (\frac{a}{1-x} + d + \hat \mu )^{-n},
\label{NJL+gen_int}
\end{equation}
where bound-state amplitudes determined by one gluon exchange correspond to  \mbox{$ n = 2$}.  The quantities $a, b, c, d $ depend on the mass-dimensioned scales in the system and are nonsingular in $x$: $a$ scales with the square of the spectator quark mass and other details are immaterial.   A change of variable to \mbox{$ \bar \mu =  (1-x) \hat \mu /a$} shows that \mbox{$u_K(x) \propto $}  \mbox{$[(1-x)/a ]^n$} when $a/(1-x)$ is greater than any physical mass-scale in the system.  Running of the struck quark mass over a wide domain can be accommodated.  We thus have \mbox{$u_K(x) \propto $} \mbox{$ N_K\, (1-x)^2/M_s^4$} and \mbox{$u_\pi(x) \propto $} \mbox{$N_\pi \, (1-x)^2/M_u^4$}.
Note that it is the bound-state amplitudes that completely determine the exponent $\alpha$ \cite{Holt:2010vj}: if the argument of $E_{K/\pi}$ did not diverge at large-$x$, the combined scaling effect of the propagators would vanish, giving \mbox{$ \alpha = 0$}.

The above analysis applied to the ratio leads to the approximate formula  \mbox{$u_K(1)/u_\pi(1) \sim $}  \mbox{$\frac{f_\pi}{f_K} \,(M_u/M_s)^4 \sim 0.2$}, where the ratio of Bethe-Salpeter amplitude normalization constants is estimated from the experimental $f_\pi/f_K$.   This estimate is in fair accord with our full calculation in Fig.\,\ref{fig:pi_DSE_ratio}.   The NJL model with a sharp cutoff yields $(M_u/M_s)^2$~\cite{Shigetani:1993dx}.  However, in general this lacks a physical contribution from bound state amplitudes and NJL results depend sensitively upon the regularization scheme.

With this study we have unified the computation of distribution functions that arise in analyses of deep inelastic scattering with that of numerous other properties of pseudoscalar mesons, including meson-meson scattering and the successful prediction of electromagnetic elastic and transition form factors.  Our results confirm the large-$x$ behavior of distribution functions predicted by the QCD parton model; provide a good account of the $\pi$-$N$ Drell-Yan data for $u_\pi(x)$; and our parameter-free prediction for the ratio $u_K(x)/u_\pi(x)$ agrees with extant data, showing a strong environment-dependence of the $u$-quark distribution.

%
We thank N.~A.~Souchlas for supplying the Bethe-Salpeter amplitudes for the pion and kaon.
This work was supported in part by the U.\,S.\ National Science Foundation, under grant
No.\ NSF-PHY-0903991, part of which constitutes USA-Mexico collaboration funding in partnership with the Mexican agency CONACyT; the U.\,S.\ Department of Energy, Office of Nuclear Physics, contract no.~DE-AC02-06CH11357; and the Project of Knowledge Innovation Program of the Chinese Academy of Sciences, Grant No.\ KJCX2.YW.W10.



\begin{thebibliography}{99}
\bibitem{Badier:1983mj}  J.~Badier {\it et al.}  [NA3 Collaboration],
 Z.\ Phys.\  C {\bf 18}, 281 (1983).


\bibitem{Betev:1985pg}
  B.~Betev {\it et al.}  [NA10 Collaboration],
  Z.\ Phys.\  C {\bf 28}, 15 (1985).

\bibitem{Conway:1989fs}
  J.~S.~Conway {\it et al.},
  Phys.\ Rev.\  D {\bf 39}, 92 (1989).

\bibitem{Badier:1980jq}
  J.~Badier {\it et al.}  [Saclay-CERN-College de France-Ecole Poly-Orsay
                  Collaboration],
  Phys.\ Lett.\  B {\bf 93}, 354 (1980).


\bibitem{Holt:2010vj}
  R.~J.~Holt and C.~D.~Roberts,
  Rev.\ Mod.\ Phys.\  {\bf 82}, 2991 (2010).


\bibitem{Maris:2000sk}
  P.~Maris and P.~C.~Tandy,
  Phys.\ Rev.\  C {\bf 62}, 055204 (2000).

\bibitem{Bicudo:2001jq}
  P.~Bicudo \emph{et al}., 
  Phys.\ Rev.\  D {\bf 65}, 076008 (2002).

\bibitem{Roberts:2010rn}
  H.~L.~L.~Roberts \emph{et al}., 
  Phys.\ Rev.\  C {\bf 82}, 065202 (2010).


\bibitem{Best:1997qp}
  C.~Best {\it et al.},
  Phys.\ Rev.\  D {\bf 56}, 2743 (1997);
  W.~Detmold, W.~Melnitchouk and A.~W.~Thomas,
  Phys.\ Rev.\  D {\bf 68}, 034025 (2003).

\bibitem{Shigetani:1993dx}
  T.~Shigetani, K.~Suzuki and H.~Toki,
  Phys.\ Lett.\  B {\bf 308}, 383 (1993).

\bibitem{Weigel:1999pc}
  H.~Weigel, E.~Ruiz Arriola and L.~P.~Gamberg,
  Nucl.\ Phys.\  B {\bf 560}, 383 (1999);
  W.~Bentz, T.~Hama, T.~Matsuki and K.~Yazaki,
  Nucl.\ Phys.\  A {\bf 651}, 143 (1999).

\bibitem{Szczepaniak:1993uq}
  A.~Szczepaniak, C.~R.~Ji and S.~R.~Cotanch,
  Phys.\ Rev.\  D {\bf 49}, 3466 (1994);
  T.~Frederico and G.~A.~Miller,
  Phys.\ Rev.\  D {\bf 50}, 210 (1994).

\bibitem{Dorokhov:2000gu}
  A.~E.~Dorokhov and L.~Tomio,
  Phys.\ Rev.\  D {\bf 62}, 014016 (2000).

\bibitem{Farrar:1975yb}
  G.~R.~Farrar and D.~R.~Jackson,
  Phys.\ Rev.\ Lett.\  {\bf 35}, 1416 (1975).


\bibitem{Brodsky:1994kg}
  S.~J.~Brodsky, M.~Burkardt and I.~Schmidt,
  Nucl.\ Phys.\  B {\bf 441}, 197 (1995);
  X.~Ji, J.~P.~Ma and F.~Yuan,
  Phys.\ Lett.\  B {\bf 610}, 247 (2005).

\bibitem{Roberts:1994dr}
  C.~D.~Roberts and A.~G.~Williams,
  Prog.\ Part.\ Nucl.\ Phys.\  {\bf 33}, 477 (1994);
  P.~C.~Tandy,
  Prog.\ Part.\ Nucl.\ Phys.\  {\bf 39}, 117 (1997);
  C.~D.~Roberts and S.~M.~Schmidt,
  Prog.\ Part.\ Nucl.\ Phys.\  {\bf 45}, S1 (2000);
%
  P.~Maris and C.~D.~Roberts,
  Int.\ J.\ Mod.\ Phys.\  E {\bf 12}, 297 (2003);
  L.~Chang and C.~D.~Roberts,
  arXiv:1003.5006 [nucl-th].

\bibitem{Munczek:1994zz}
  H.~J.~Munczek,
  Phys.\ Rev.\  D {\bf 52} (1995) pp.~4736-4740;
%
  A.~Bender, C.~D.~Roberts and L.~Von Smekal,
  Phys.\ Lett.\  B {\bf 380}(1996) pp.~7-12.


\bibitem{Hecht:2000xa}
  M.~B.~Hecht, C.~D.~Roberts and S.~M.~Schmidt,
  Phys.\ Rev.\  C {\bf 63}, 025213 (2001).

\bibitem{Maris:1999bh}
  P.~Maris and P.~C.~Tandy,
  Phys.\ Rev.\  C {\bf 61}, 045202 (2000).

\bibitem{Maris:2002mz}
  P.~Maris and P.~C.~Tandy,
  Phys.\ Rev.\  C {\bf 65}, 045211 (2002).

\bibitem{Holl:2005vu}
  A.~H\"oll, A.~Krassnigg, P.~Maris, C.~D.~Roberts and S.~V.~Wright,
  Phys.\ Rev.\  C {\bf 71}, 065204 (2005).

\bibitem{Jaffe:1983hp}
  R.~L.~Jaffe,
  Nucl.\ Phys.\  B {\bf 229}, 205 (1983);
  ``Deep Inelastic Scattering With Application To Nuclear Targets,'' in \emph{Proceedings of the Los Alamos School on Quark Nuclear Physics}, Los Alamos, NM, Jun 10-14, 1985, eds. M.~B.~Johnson and A.~Picklesimer (Wiley, New York, 1986).

\bibitem{Ellis:1991qj}
  R.~K.~Ellis, W.~J.~Stirling and B.~R.~Webber,
  Camb.\ Monogr.\ Part.\ Phys.\ Nucl.\ Phys.\ Cosmol.\  {\bf 8}, 1 (1996).

\bibitem{Maris:1997hd}
  P.~Maris, C.~D.~Roberts and P.~C.~Tandy,
  Phys.\ Lett.\  B {\bf 420}, 267 (1998).

\bibitem{Maris:1997tm}
  P.~Maris and C.~D.~Roberts,
  Phys.\ Rev.\  C {\bf 56}, 3369 (1997).

\bibitem{Maris:1999nt}
  P.~Maris and P.~C.~Tandy,
  Phys.\ Rev.\  C {\bf 60}, 055214 (1999).

\bibitem{Chang:2009zb}
  L.~Chang and C.~D.~Roberts,
  Phys.\ Rev.\ Lett.\  {\bf 103}, 081601 (2009).

\bibitem{Roberts:1994hh}
  C.~D.~Roberts,
  Nucl.\ Phys.\  A {\bf 605}, 475 (1996).

\bibitem{PDG10}
  K.~Nakamura {\it et al.}  [Particle Data Group],
  J.\ Phys.\ G {\bf 37} (2010) 075021.

\bibitem{Nguyen_PhD10} T.~T.~Nguyen, PhD dissertation, Kent State
  University.

\bibitem{Wijesooriya:2005ir}
  K.~Wijesooriya, P.~E.~Reimer and R.~J.~Holt,
  Phys.\ Rev.\  C {\bf 72}, 065203 (2005).

\bibitem{Aicher:2010cb}
  M.~Aicher, A.~Schafer and W.~Vogelsang,
  Phys.\ Rev.\ Lett.\  {\bf 105}, 252003 (2010).


\bibitem{Sutton:1991ay}
  P.~J.~Sutton, A.~D.~Martin, R.~G.~Roberts and W.~J.~Stirling,
  Phys.\ Rev.\  D {\bf 45}, 2349 (1992).

\bibitem{Chang:2010xs}
  L.~Chang and C.~D.~Roberts,
  AIP Conf.\ Proc.\  {\bf 1261}, 25 (2010).

\end{thebibliography}


\end{document}